\documentclass[twocolumn,prb]{revtex4}% Physical Review Letter
\usepackage{graphicx}    % Include figure files
\usepackage{dcolumn}     % Align table columns on decimal point
\usepackage{bm}          % bold math
\usepackage{amsmath}
\begin{document}
\preprint{qmctriplet}

%--- title ---
\title{Competition between singlet and triplet pairings in 
Na$_x$ CoO$_2 \cdot y$ H$_2$O
}
%--- author ---
\author{
Kazuhiko Kuroki
}
\affiliation{
Department of Applied Physics and
Chemistry, The University of Electro-Communications,
Chofu, Tokyo 182-8585, Japan}

\author{Yukio Tanaka}
\affiliation{Department of Applied Physics,
Nagoya University, Nagoya, 464-8603, Japan}

\author{Ryotaro Arita}
\affiliation{Department of Physics,
University of Tokyo, Hongo 7-3-1, Tokyo 113-0033, Japan}

\date{\today}% It is always \today, today,
%   Abstract
%-----------------------------------------------------------
\begin{abstract}
We discuss the 
pairing symmetry of a cobaltate superconductor 
Na$_x$CoO$_2\cdot y$ H$_2$O by adopting an effective single band model
that takes into account the $e_g'$ hole pockets, as discussed 
in our previous paper [to appear in Phys. Rev. Lett.] 
Here we consider the off-site repulsions in addition to the 
on-site repulsion considered in our previous study.
We show that the spin-triplet $f$-wave pairing proposed in our 
previous study is robust to some 
extent even in the presence of off-site repulsions. However, 
$f$-wave pairing gives way to singlet pairings for sufficiently 
large values of off-site repulsions. Among the singlet pairings, 
$i$-wave and extended $s$-wave pairings are 
good candidates which do not 
break time reversal symmetry below $T_c$ in agreement with the 
experiments.
\end{abstract}
%-----------------------------------------------------------
\pacs{PACS numbers: }
\maketitle
%-----------------------------------------------------------
\section{Introduction}
%----------------------------------------------------------
Discovery of superconductivity in a hydrated sodium cobaltate 
Na$_x$CoO$_2\cdot y$H$_2$O \cite{Takada} has attracted much attention 
recently. 
This material is of special interest since 
it is a $3d$-electron system having a quasi-two-dimensional 
lattice structure, which resembles the high $T_c$ cuprates.
On the other hand,  it differs from the cuprates in that 
the Co ions form a triangular lattice, and also there are several
bands intersecting the Fermi level, all of which are away 
from half-filling. Another interesting point concerning this material is 
that some experiments suggest possible occurrence of an 
unconventional superconductivity.\cite{Sakurai,Waki,Fujimoto,Ishida} 
In particular, there has been a debate as to whether the 
pairing occurs in the spin singlet channel or in the triplet 
channel. An unchanged 
Knight shift across $T_c$ found in some experiments 
suggests spin-triplet pairing,\cite{Waki} while 
the Knight shift does decrease below $T_c$ in other experiments
at low magnetic field.\cite{Kobayashi,Michioka}
As for the occurrence of broken time reversal symmetry,
several $\mu SR$ experiments show that such a possibility is rather 
small.\cite{Higemoto,Uemura}

\begin{figure}[htb]
\begin{center}
\scalebox{0.8}{
\includegraphics[width=10cm,clip]{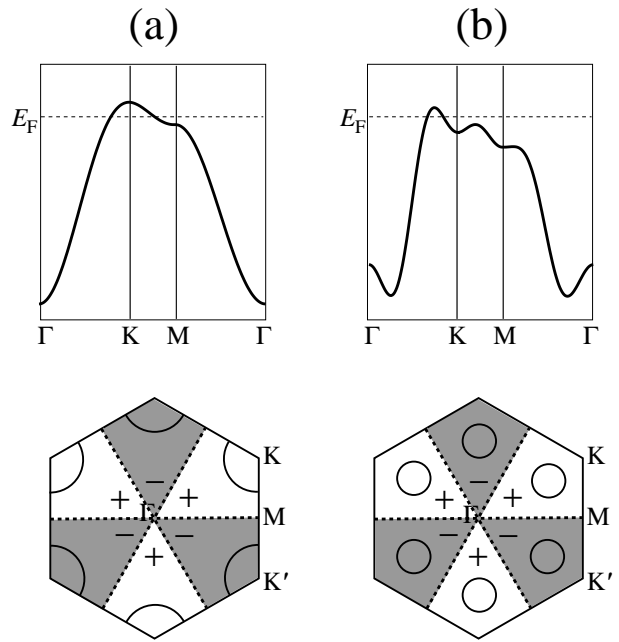}}
\caption{
Band dispersion (top) along with the Fermi surfaces (schematic)
for a small number of holes and the $f$-wave gap 
(bottom) are shown for a tight binding model on a triangular lattice 
with (a) only nearest neighbor hopping, and (b) when fourth nearest neighbor
hopping $t_4=0.2$ is introduced. 
$+-$ denote the sign of the gap function, while the 
dashed lines represent the nodal lines.
\label{fig1}}
\end{center}
\end{figure}

There has also been a singlet-triplet debate theoretically. 
Some theories propose singlet pairing, 
\cite{Baskaran,Kumar,Wang,Ogata}, others 
triplet pairing mechanisms.\cite{Tanaka,Ikeda,TanaOgata,KYTA,MYO,YMO} 
Naively, presence of 
possible ferromagnetic spin fluctuations reported in some experiments
\cite{Ishida,Kobayashi} may support 
spin-fluctuation-mediated triplet pairing scenario.
Couple of years ago, however, two of the present authors showed that 
$T_c$ of spin triplet superconductivity due to ferromagnetic 
spin fluctuations, if any, should in general be very low.\cite{AKA}
A similar conclusion has been reached by Monthoux and Lonzarich.\cite{ML}
This is because the triplet pairing interaction mediated by spin fluctuations 
is proportional to 
$\chi/2$ (in contrast to $3\chi/2$ for the singlet case), 
where $\chi$ is the spin susceptibility, while the 
effective interaction 
that enters the normal self energy is proportional 
to $3\chi/2$, so that the suppression of pairing due to the normal self-energy 
overpowers the pairing effect.
We then showed that 
this difficulty for spin-fluctuation-mediated 
triplet pairing may be eased in systems having 
`disconnected Fermi surfaces', where the nodal lines of the gap 
can run {\it in between} the Fermi surfaces to open up a 
full gap {\it on the Fermi surfaces}.\cite{KA} 
Since pairing with high angular momentum is degraded by the presence of 
gap nodes on the Fermi surfaces, opening a full gap 
may result in an enhanced pairing.\cite{KA,KA2,TKA,KTA,TKA2,OKAA}
As an example, we considered 
the Hubbard model on a triangular lattice, where the Fermi surface
becomes disconnected into two pieces centered around the K point and the
K$'$ point (see Fig.\ref{fig1}(a)) 
when there are a small number of holes (or 
electrons, depending on the sign of the hopping integral).
Using fluctuation exchange (FLEX) approximation\cite{Bickers,Dahm,Grabowski} 
and solving the linearized 
{\'E}liashberg equation, we have shown that a finite $T_c$ 
can be obtained for spin-triplet 
$f$-wave pairing, where the nodes of the gap run between the 
Fermi surfaces.\cite{KA} 
We have also confirmed this conclusion\cite{AKA2} using 
dynamical cluster approximation,\cite{Jarrell} which is a 
non-perturbative approach.
%
%at 
%a $T_c$ of order $0.001t$ ($t$ is the hopping integral), 
%which is much higher than the $T_c$ obtained by the same method, 
%if any, for systems having enhanced ferromagnetic
%spin fluctuations but with conncted Fermi surfaces.

\begin{figure}[htb]
\begin{center}
\scalebox{0.8}{
\includegraphics[width=10cm,clip]{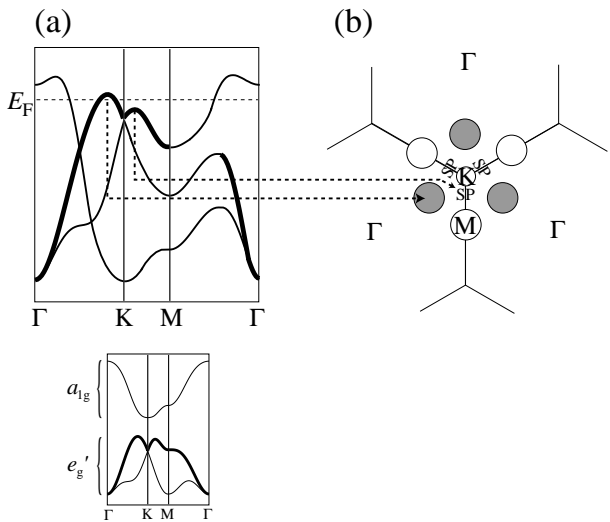}}
\caption{
(a) A schematic plot of the band structure of Na$_x$CoO$_2$ following
the calculation results of ref.\onlinecite{Singh}.
The thick line denotes the portion of the upper part of the bands 
which has $e_g'$ character to some extent. Bottom inset:
dispersion of a three band tight binding model, where large 
$a_{1g}$-$e_g'$ level offset as well as third nearest neighbor hoppings 
between same orbitals is introduced in addition to the hoppings considered in 
ref.\cite{Koshibae}.
(b) Brillouin zone is shown in the extended zone scheme. 
The points denoted as SP are saddle points since the energy of the thick band 
at SP is lower than those around the gray circles, while it is 
higher than those around the open circles. 
\label{fig2}}
\end{center}
\end{figure}

In a recent study,\cite{KYTA} we have looked at the  band calculation 
results of Na$_{x}$CoO$_2$\cite{Singh} 
from this point of view, where we find that the pocket-like 
Fermi surfaces (hole pockets)
near the K and K$'$ points (as in the bottom figure of 
Fig.\ref{fig1}(b)), originating from the band having $e_g'$ character
to some extent in the notation of ref.\onlinecite{Singh} 
(denoted by the thick line in Fig.\ref{fig2}(a)), 
are disconnected in a similar sense as in 
the triangular lattice, namely the nodes of the $f$-wave gap 
do not intersect the Fermi surfaces 
(although they do intersect the large Fermi surface
around the $\Gamma$ point originating from the bands 
having $a_{1g}$ character). 

In fact, a strong motivation 
to focus on the $e_g'$ hole pockets is the 
presence of van Hove singularity (vHS) near the K point.
Namely, as can be seen from the band structure 
shown in Fig.\ref{fig2}, there exist saddle points at points denoted as SP,
where the density of states (DOS) takes a large value. Since this large 
DOS lies close to the Fermi level\cite{Singh}, 
it is likely that the band structure around 
the K and K$'$ points strongly affects the low energy properties 
of this material. In particular, ferromagnetic 
spin fluctuations may arise due to this high DOS near the Fermi level.

In ref.\onlinecite{KYTA}, 
we have adopted a model where we separate out the portion of the 
bands which has $e_g'$ character to some extent,  
and in particular focus only on the upper $e_g'$ band,
which contributes to the formation of the pocket Fermi surfaces 
and to the large DOS at the vHS, while neglecting 
the lower $e_g'$ band, which has only small DOS 
near the Fermi level due to the linear dispersion at the band top. 
Although this portion of the band, 
shown by the thick curves in Fig.\ref{fig2}(a),
is disconnected between M and $\Gamma$ points due to $a_{1g}$-$e_g'$
hybridization, it originally comes from a single band having 
$e_g'$ character, as can be understood from the 
inset of Fig.\ref{fig2}(a), where a tight binding band dispersion is given 
for a case when large $a_{1g}$-$e_g'$ level offset is introduced.
We have found that the thick portion of the band 
in Fig.\ref{fig2}(a) (apart from the 
missing part between M and $\Gamma$ points) 
can be roughly reproduced by a single band tightbinding 
model on a triangular lattice 
with hopping integrals up to fourth nearest neighbors.
Namely, the dispersion of the tight binding model 
on an isotropic triangular lattice with only nearest neighbor hoppings 
takes its maximum at the K point (Fig.\ref{fig1}(a)), while 
the band top moves towards the $\Gamma$ point 
when the fourth neighbor hopping is introduced  (Fig.\ref{fig1}(b)),
resulting in a pocket like Fermi surface that lies between the 
K and $\Gamma$ points when small amount of holes are present.
Moreover, the band structure around the K point resembles 
that of the actual material (compare Fig.\ref{fig1}(b) and the 
thick curve in Fig.\ref{fig2}), so the vHS due to the saddle points SP in 
Fig.\ref{fig2} is also reproduced by this effective model.
As for electron-electron interactions, we considered a model with 
only the on-site repulsion, where 
we found using FLEX 
that spin triplet $f$-wave pairing indeed dominates and has a 
finite $T_c$.\cite{KYTA} 

In the present study, we continue to study the 
single band model having 
hoppings up to fourth nearest neighbors, but this time with repulsive 
interactions up to second nearest neighbors 
to investigate the effect of the off-site repulsions.
We find that (i) spin triplet $f$-wave pairing is robust to some extent 
even in the presence of off-site repulsions, but (ii) for sufficiently 
large off-site repulsions, singlet pairings can dominate over triplet 
pairings.


\section{Formulation}
\subsection{Model}
In standard notations, the Hamiltonian considered in the present study 
is given in the form,
\begin{align}
H=\sum_{{ij},\sigma}\left(t_{ij}c^\dagger_{i\sigma}c_{j\sigma}+{\rm h.c.}
\right)
+U\sum_i n_{i\uparrow} n_{i\downarrow}
\nonumber\\
+V_1\sum_{\langle ij\rangle',\sigma,\sigma'} n_{i\sigma} n_{j\sigma'}+
V_2\sum_{\langle ij\rangle'',\sigma,\sigma'} n_{i\sigma} n_{j\sigma'}
\end{align}
where $t_{ij}=t_1$, $t_2$, $t_3$, and $t_4$ are the nearest, 
second nearest, third nearest and fourth nearest neighbor hopping integrals, 
respectively,\cite{comment} 
and $U$, $V_1$, $V_2$, represent on-site, nearest neighbor,
second nearest neighbor electron-electron repulsions, respectively.
The summations $\sum_{\langle ij\rangle'}$ and 
$\sum_{\langle ij\rangle''}$ are taken over pairs of nearest and 
next nearest neighbors, respectively.
We define the hole density as $n_h=2-n$, where $n$ is the 
electron band filling (number of electrons/number of sites).
$n_h$ is restricted to a small number in the present study 
because the $e_g'$ hole pockets 
are small. We take $-t_1=1$ as the unit of the energy throughout the study.

\subsection{Method}
To make the calculation feasible at low temperatures even 
in the presence of off-site repulsions up to second nearest neighbors, 
here we adopt a random phase approximation (RPA) approach.
The singlet ($V^s$) and triplet ($V^t$) pairing interactions 
are given within RPA as follows.
\begin{align}
\label{1}
V^{s}(\bm{q})=
U + V({\bm q}) + \frac{3}{2}U^{2}\chi_{s}(\bm{q})
\nonumber\\
-\frac{1}{2}(U + 2V({\bm q}) )^{2}\chi_{c}(\bm{q})
\end{align}
\begin{align}
\label{2}
V^{t}(\bm{q})=
V({\bm q}) - \frac{1}{2}U^{2}\chi_{s}(\bm{q})
\nonumber\\
-\frac{1}{2}(U + 2V({\bm q}) )^{2}\chi_{c}(\bm{q})
\end{align}
Here, $V(\bm{q})$ is the Fourier transform of the off-site repulsions.
$\chi_{s}$ and $\chi_{c}$ are the spin and charge 
susceptibilities, respectively,  which are given as 
\begin{align}
\label{4}
\chi_{s}(\bm{q},\omega_{l})=\frac{\chi_{0}(\bm{q})}
{1 - U\chi_{0}(\bm{q})}
\nonumber\\
\chi_{c}(\bm{q})=\frac{\chi_{0}(\bm{q})}
{1 + (U + 2V(\bm{q}) )\chi_{0}(\bm{q})}.
\end{align}
Here $\chi_{0}$ is the bare susceptibility given by 
\[
\chi_{0}(\bm{q})
=\frac{1}{N}\sum_{\bm{p}} 
\frac{ f(\varepsilon_{\bm{p +q}})-f(\varepsilon_{\bm{p}}) }
{\varepsilon_{\bm{p }} -\varepsilon_{\bm{p+q}}}
\]
with
$\varepsilon_{\bm{k}}$ being the energy dispersion measured from the 
chemical potential and $f(\varepsilon_{\bm{p}})$ is the Fermi 
distribution function. We take 128$\times$128 $k$-point meshes in 
the actual numerical calculations.

To obtain the onset of the superconducting state, 
we solve the gap equation within the 
weak-coupling theory, 
\begin{equation}
\lambda \Delta(\bm{k})
=-\sum_{\bm{k'}} V^{s,t}(\bm{k-k'})
\frac{ \rm{tanh}(\beta \varepsilon_{{\bm{k'} }}/2) }{2 \varepsilon_{\bm{k'}} }
\Delta(\bm{k'}).
\end{equation}
The transition temperature $T_{C}$ is determined by the condition, 
$\lambda=1$. 
In the present approach, $\omega$ dependence 
as well as the self energy correction is neglected.
Although this approximation is quantitatively insufficient, 
we believe that 
the present approach suffices for the purpose of discussing the 
{\it competition} (relative tendency toward pairing) 
among various pairing symmetries.

\subsection{Possible pairing symmetries}
\begin{figure}[htb]
\begin{center}
\scalebox{0.8}{
\includegraphics[width=10cm,clip]{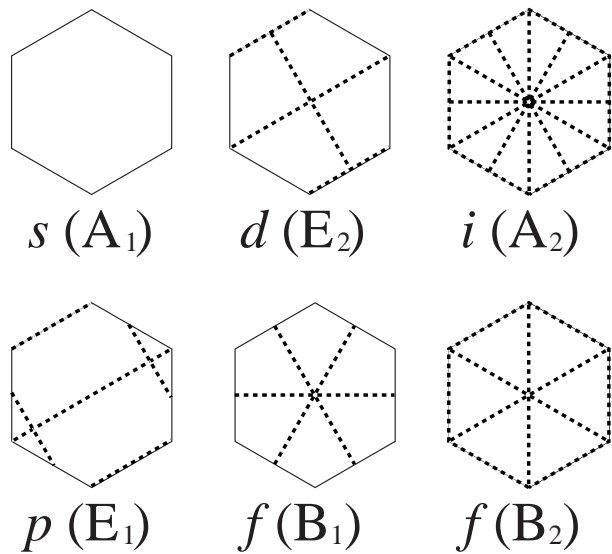}}
\caption{
Possible pairing symmetries of the present model classified 
according to the irreducible representations of D$_6$. 
The dashed lines represent the basic nodal lines.
\label{fig3}}
\end{center}
\end{figure}

Pairing symmetries are classified according to the irreducible 
representations of the point group D$_6$, as shown in Fig.\ref{fig3}.
For the cases we considered, the $f$-wave pairing having 
$B_2$ symmetry turned out to have small values of $\lambda$,
so in the following sections we consider $s$, $p$, $d$, $f$, $i$-wave 
pairings having $A_1$, $E_1$, $E_2$, $B_1$, $A_2$ symmetry, 
respectively.

\section{Results}

\subsection{Case without off-site repulsions}
\label{nooffsite}
\begin{figure}[htb]
\begin{center}
\scalebox{0.8}{
\includegraphics[width=10cm,clip]{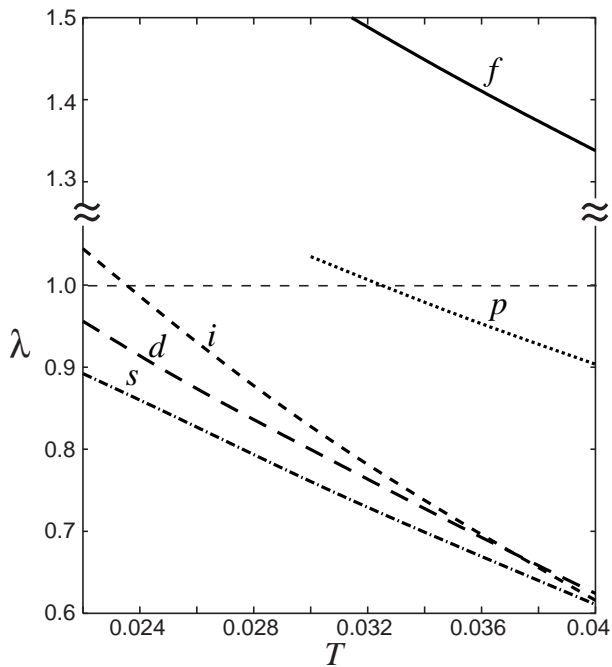}}
\caption{
The eigenvalue of the gap equation plotted as a function of temperature
for each pairing symmetry. $t_2=t_3=0$, $t_4=0.18$, 
$U=2.4$, $V_1=V_2=0$, $n_h=0.18$. 
\label{fig4}}
\end{center}
\end{figure}

\begin{figure}[htb]
\begin{center}
\scalebox{0.8}{
\includegraphics[width=10cm,clip]{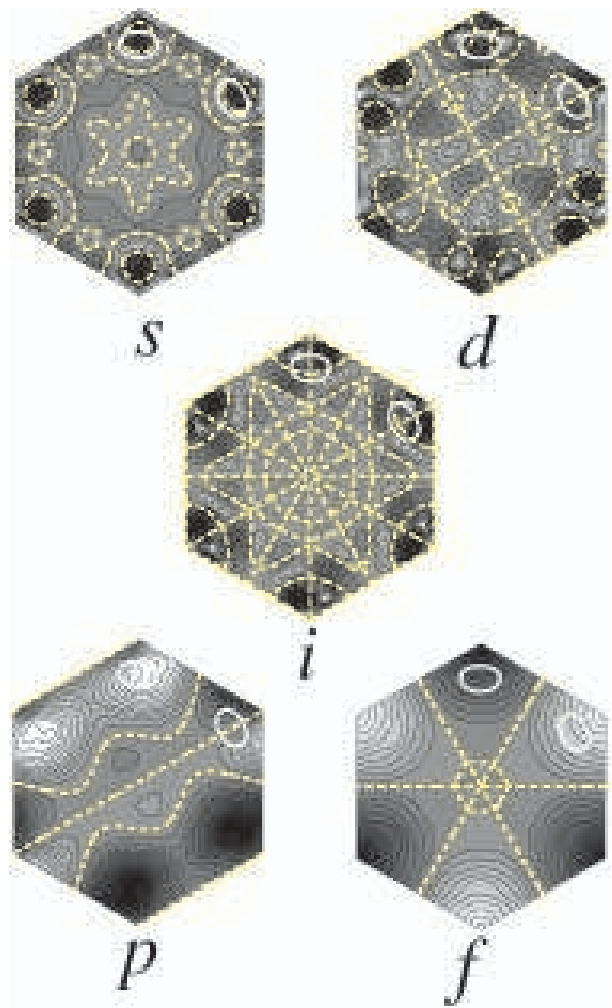}}
\caption{
Contour plot of the gap functions for each pairing symmetry. 
The dashed lines represent the nodes of the gap, and the 
white rings are the hole pockets (only two out of the six are shown
for clarity). Parameter values are the same as in Fig.\ref{fig4} 
and $T=0.03$. 
All the gap functions are plotted at $T=0.03$ throughout the study.
\label{fig5}}
\end{center}
\end{figure}

We first show results for a case without off-site repulsions,
where we take $t_2=t_3=0$, $t_4=0.18$, $n_h=0.18$, $U=2.4$ and $V_1=V_2=0$. 
As seen in Fig.\ref{fig4}, the $f$-wave pairing strongly 
dominates over other pairing symmetries, which is consistent 
with our previous study.\cite{KYTA} 
The reason for this is: (i) the spin fluctuations are 
nearly ferromagnetic (see Fig.\ref{fig8}(a)), 
and (ii) the nodes of the $f$-wave gap do not 
intersect the Fermi surfaces as seen in Fig.\ref{fig5}.

\subsection{Case with moderate nearest neighbor repulsion}
\label{moderate}
\begin{figure}[htb]
\begin{center}
\scalebox{0.8}{
\includegraphics[width=10cm,clip]{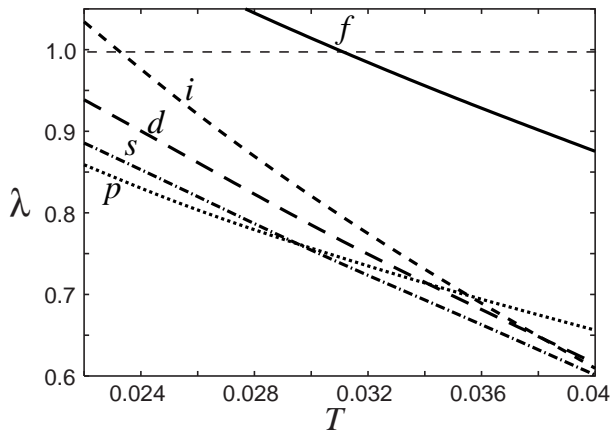}}
\caption{
A plot similar to Fig.\ref{fig4} with $V_1=0.7$, $V_2=0.4$, and 
other parameter values taken to be the same.
\label{fig6}}
\end{center}
\end{figure}

\begin{figure}[htb]
\begin{center}
\scalebox{0.8}{
\includegraphics[width=10cm,clip]{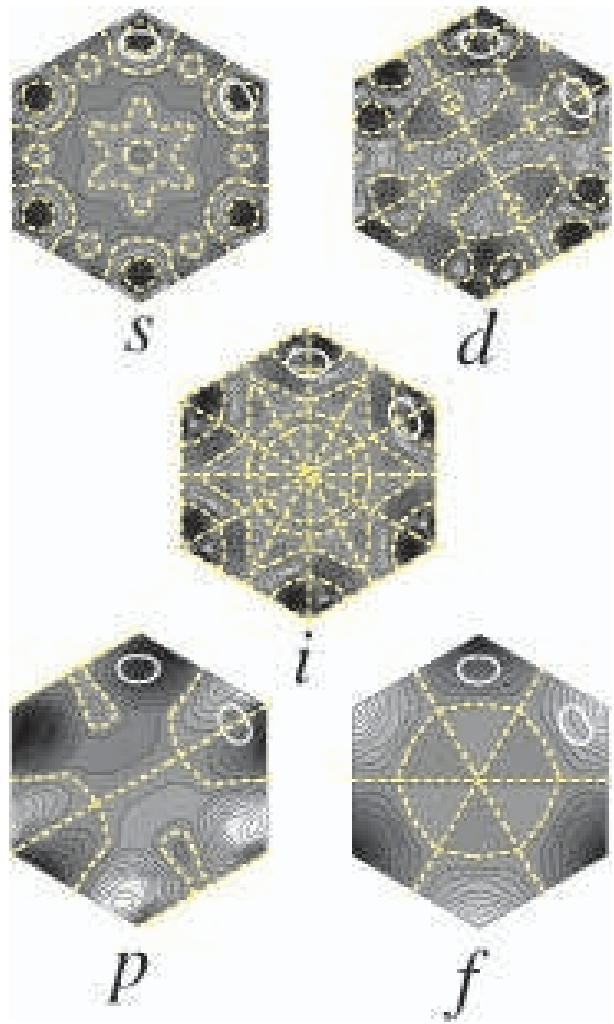}}
\caption{
Gap functions for the parameter values of Fig.\ref{fig6}. 
\label{fig7}}
\end{center}
\end{figure}

\begin{figure}[htb]
\begin{center}
\scalebox{0.8}{
\includegraphics[width=10cm,clip]{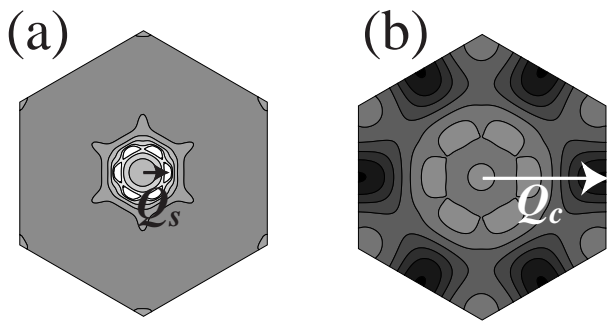}}
\caption{
Contour plots of the spin (a) and the charge (b) part of the 
pairing interactions for the parameter values 
of Fig.\protect\ref{fig6}. The spin part takes its maximum 
value $\sim 25$ at a short wave but finite vector $\protect\bm{Q}_s$, 
while the charge part takes its minimum value $\sim -3$
at the M point.
\label{fig8}}
\end{center}
\end{figure}

\begin{figure}[htb]
\begin{center}
\scalebox{0.8}{
\includegraphics[width=7cm,clip]{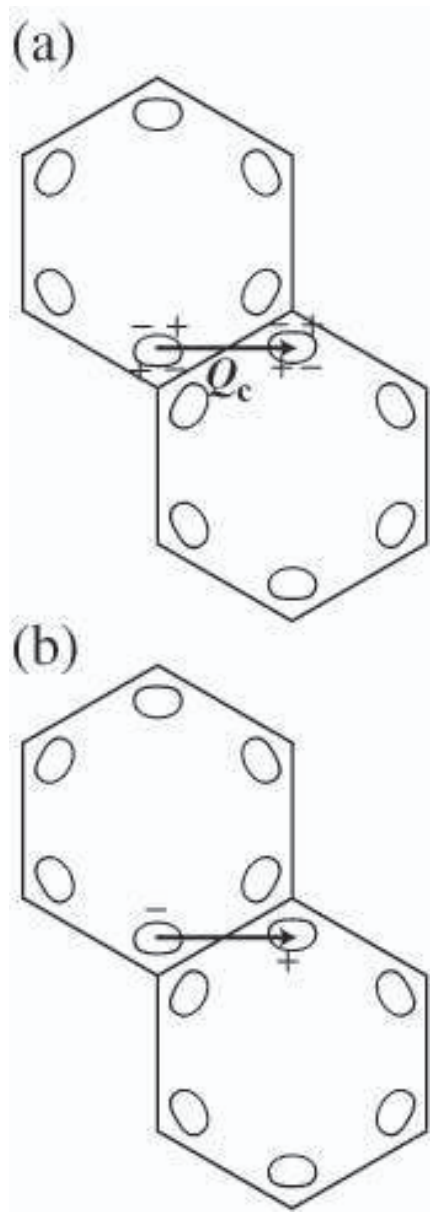}}
\caption{
The reason why singlet pairings (here we take 
the $i$-wave gap in Fig.\ref{fig7} as an example) 
are barely affected by off-site
repulsions (a), while $f$-wave pairing is suppressed (b).
$Q_c$ is the wave vector at which the charge part of the pairing 
interaction takes its maximum absolute value.
\label{fig9}}
\end{center}
\end{figure}

We now turn on the off-site repulsions.
As a case where a moderate nearest neighbor repulsion exists,
we take $V_1=0.7$ and $V_2=0.4$, while other parameter values 
are taken to be the same as in section \ref{nooffsite}.
We see in Fig.\ref{fig6} that the $f$-wave pairing 
still dominates over the others, but it is suppressed compared to the 
case without off-site repulsions. 
On the other hand, $\lambda$ of singlet $i$,$d$, and $s$-wave pairings, 
along with the singlet gap functions shown in Fig.\ref{fig7}, 
are barely affected by the off-site repulsions. 
To understand the reason why $f$-wave is suppressed while
singlet pairings are unaffected by off-site repulsions, 
we show in Fig.\ref{fig8} the 
pairing interaction in momentum space.
Here we define the ``spin part'' and the ``charge part'' 
of the pairing interaction as $U^2\chi_s({\bm q})$ and 
$V(\bm{q})-\frac{1}{2}(U+2V({\bf q}))^2\chi_c(\bm{q})$, respectively.
The spin part peaks near the $\Gamma$ point, while the 
charge part takes a broad negative minimum value around the M point.
Thus, the effect of the pair scattering between two pockets that is 
mediated mainly by the charge part cancels out when there are nodes 
on the pockets (Fig.\ref{fig9}(a)), while  
the sign change of the $f$-wave gap 
is unfavorable in the presence 
of the charge part, which has a negative sign (Fig.\ref{fig9}(b)).

\subsection{Case with large nearest neighbor repulsion}
\label{large}
\begin{figure}[htb]
\begin{center}
\scalebox{0.8}{
\includegraphics[width=10cm,clip]{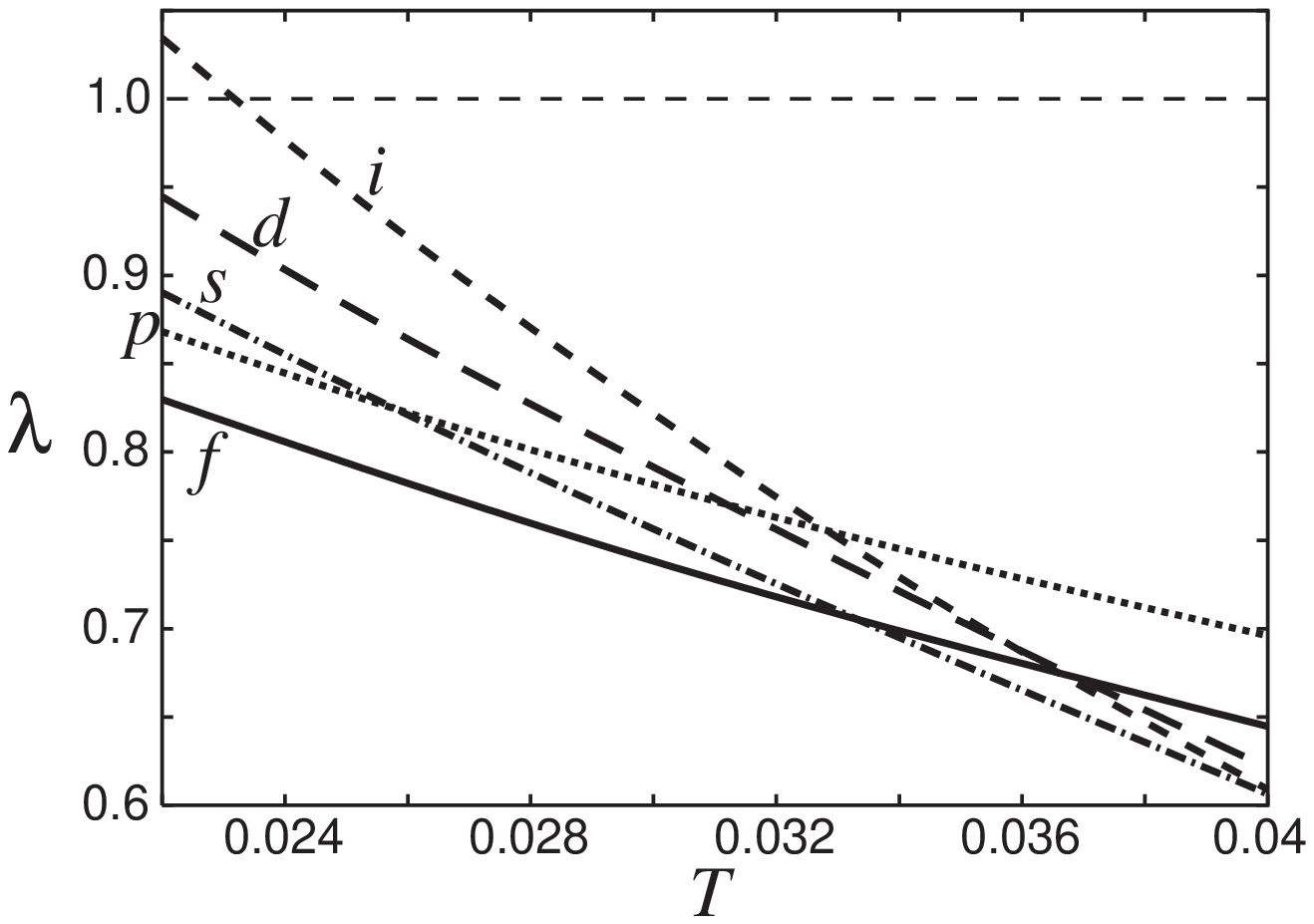}}
\caption{
A plot similar to Fig.\ref{fig6} with $V_1=1.0$ and other 
parameter values taken to be the same.
\label{fig10}}
\end{center}
\end{figure}

\begin{figure}[htb]
\begin{center}
\scalebox{0.8}{
\includegraphics[width=10cm,clip]{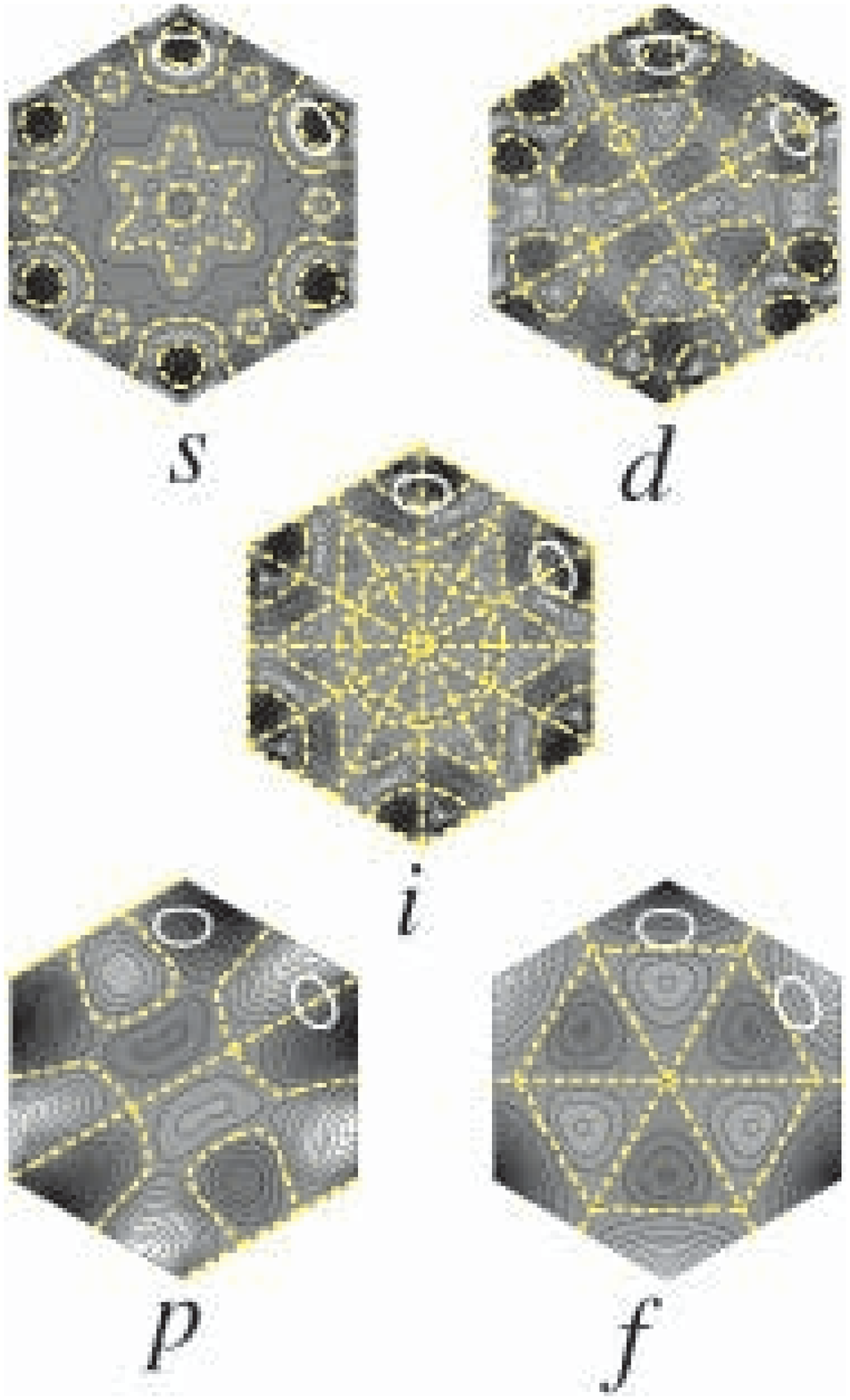}}
\caption{
Gap functions for the parameter values of Fig.\ref{fig10}.
\label{fig11}}
\end{center}
\end{figure}

For a larger nearest neighbor repulsion, where we take
the parameters values to be the same as in 
section \ref{moderate} except that $V_1=1.0$, 
we find that $f$-wave pairing gives way to 
singlet pairings, as shown in Fig.\ref{fig10}. 
Among the singlet pairings, which are found to closely 
compete in a wide range of parameter values, $i$-wave 
pairing surprisingly has the largest $T_c$ for the present set of 
parameter values. Note that 
$f$-wave is no longer full-gaped on the Fermi surfaces 
due to the additional nodes 
as seen in Fig.\ref{fig11}. 


\section{Discussion}
\subsection{Intuitive picture in real space}
The reason why triplet pairings are suppressed by off-site repulsions
while singlet pairings are barely affected can 
intuitively be understood in a real space picture. As seen in 
Fig.\ref{fig5} and Fig.\ref{fig7}, 
the triplet pairing gap functions have relatively small number of 
nodes, which shows that these pairing takes place at short distances 
in real space. In particular, $f$-wave gap in the absence of off-site 
repulsions (Fig.\ref{fig5}) essentially has the form 
$\sin(\frac{\sqrt{3}}{2}k_x+\frac{1}{2}k_y)-\sin(k_y)+
\sin(-\frac{\sqrt{3}}{2}k_x+\frac{1}{2}k_y)$, which can be 
rewritten, apart from a constant, as 
\begin{eqnarray}
&&\exp(i{\bm k}\cdot{\bm a})
-\exp(i{\bm k}\cdot{\bm b})
+\exp(i{\bm k}\cdot{\bm c})\nonumber\\
&-&\exp(-i{\bm k}\cdot{\bm a})
+\exp(-i{\bm k}\cdot{\bm b})
-\exp(-i{\bm k}\cdot{\bm c}),
\end{eqnarray}
where ${\bm a}=(\frac{\sqrt{3}}{2},\frac{1}{2})$, 
${\bm b}=(0,1)$, ${\bm c}=(-\frac{\sqrt{3}}{2},\frac{1}{2})$, 
and ${\bm k}=(k_x,k_y)$.

Since $\pm{\bm a}$, $\pm{\bm b}$, and $\pm{\bm c}$ are the 
coordinates (in units of the lattice constant) of the six nearest neighbors,
the $f$-wave pairing in the absence of off-site repulsion is 
essentially a nearest neighbor odd-parity pairing.
By contrast, the singlet pairings have many additional 
nodal structures compared to the 
basic nodal structure shown in Fig.\ref{fig3}. This means that the 
pairs are mainly formed at large distances in real space.
In fact, the result that $\lambda$ is unaffected by $V_1$ and $V_2$ 
is consistent with the picture 
that the singlet pairs are formed at distances farther than
second nearest neighbors. 

\begin{figure}[htb]
\begin{center}
\scalebox{0.8}{
\includegraphics[width=10cm,clip]{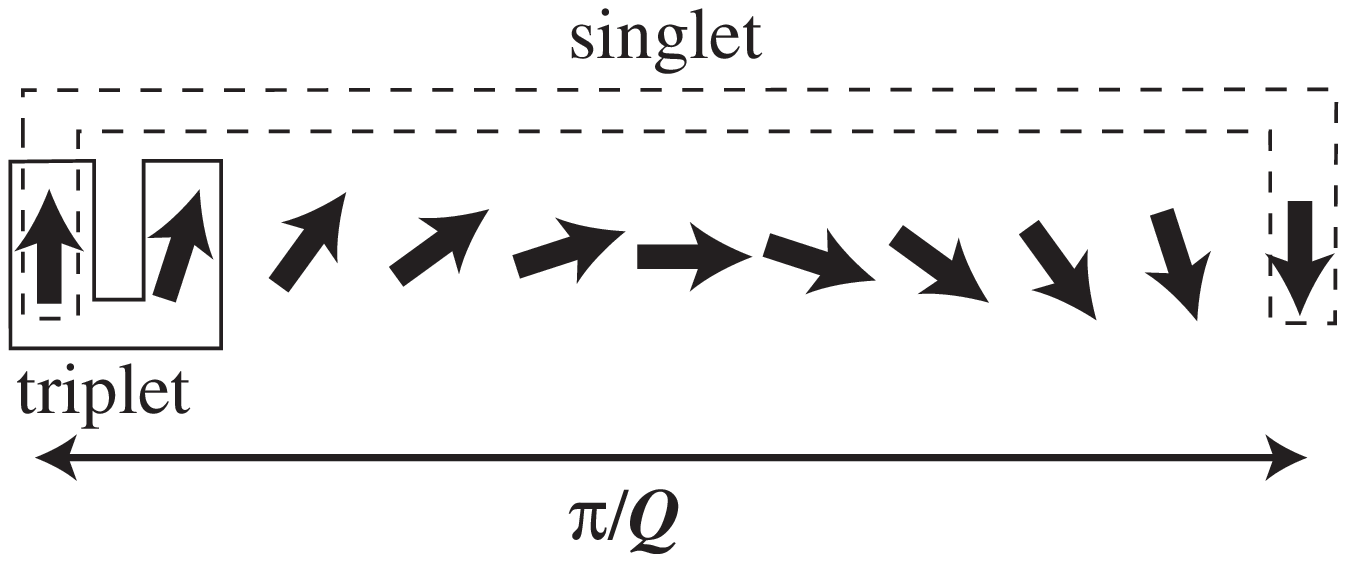}}
\caption{
An intuitive real space picture of singlet and triplet pairings 
when nearly ferromagnetic spin fluctuations are present.
\label{fig12}}
\end{center}
\end{figure}

The reason why singlet pairs are formed at large distances while
triplet pairs are formed at short distances can intuitively be 
understood from a nearly ferromagnetic spin alignment in 
real space as shown in Fig.\ref{fig12}. Namely, electrons 
at close distances have nearly parallel spins, while 
the electrons can have antiparallel spins 
only when they are separated far away. 

\begin{figure}[htb]
\begin{center}
\scalebox{0.8}{
\includegraphics[width=10cm,clip]{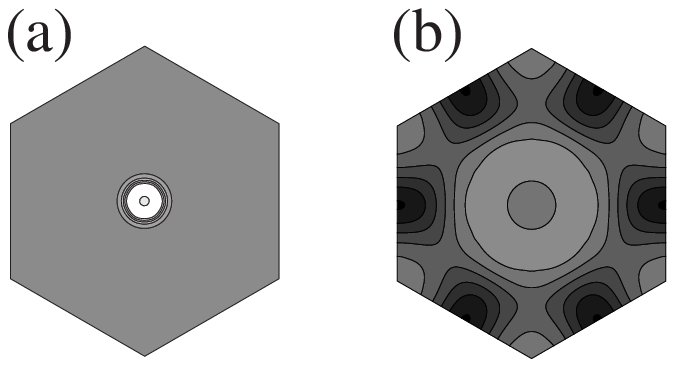}}
\caption{
The spin (a) and charge (b) part of the pairing interaction 
for $t_2=t_3=0$, $t_4=-0.15$, $U=2.4$, $V_1=1.0$, $V_2=0.4$, $n_n=0.1$.
The spin part takes its maximum 
value $\sim 35$ at a vector $\protect\bm{Q}_s\sim \protect\bm{0}$, 
while the charge part takes its minimum value $\sim -3$
at the M point.
\label{fig13}}
\end{center}
\end{figure}

In order to reinforce this real space intuitive picture,  
we look into a case where the spin susceptibility 
peaks at positions more closer to the $\Gamma$ point, 
namely when the spin structure is more purely ferromagnetic than in
the cases considered in sections \ref{moderate} and \ref{large}.
We take $t_4=0.15$ and $n_h=0.1$, which moves the vHS and the 
Fermi level towards the band edge, thus making the spin structure 
more ferromagnetic as seen in Fig.\ref{fig13}(a).
\begin{figure}[htb]
\begin{center}
\scalebox{0.8}{
\includegraphics[width=10cm,clip]{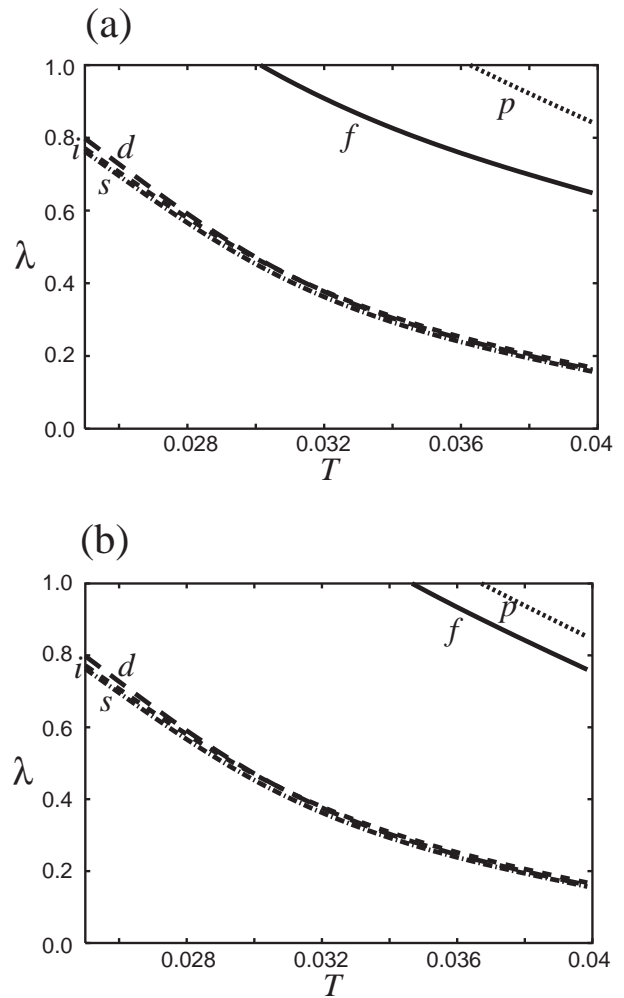}}
\caption{
Plots similar to Fig.\ref{fig6} for 
$t_2=t_3=0$, $t_4=0.15$, $U=2.4$, $V_2=0.4$, $n_n=0.1$. 
$V_1=1.0$ (a) and $V_1=0.7$(b).
\label{fig14}}
\end{center}
\end{figure}
\begin{figure}[htb]
\begin{center}
\scalebox{0.8}{
\includegraphics[width=10cm,clip]{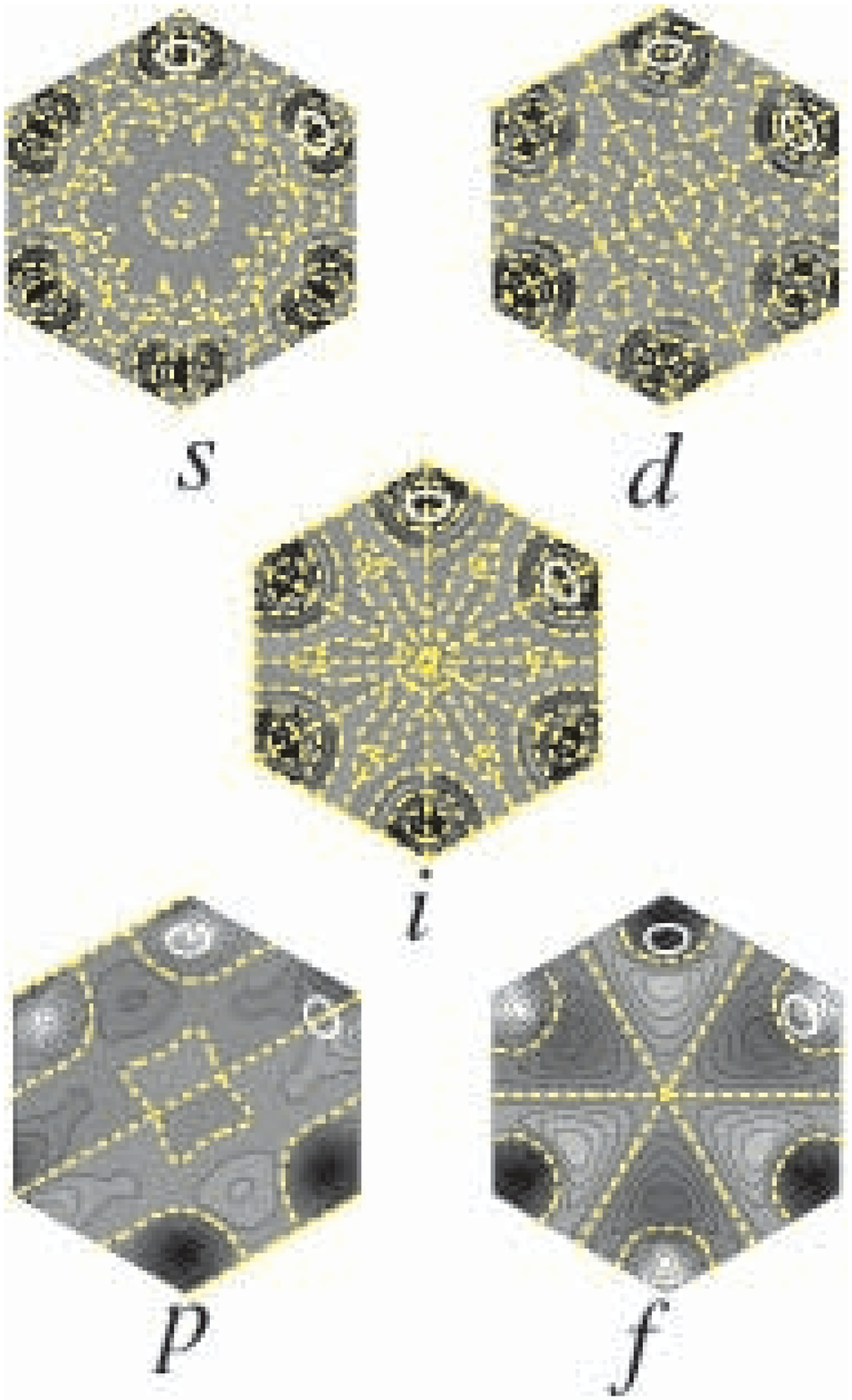}}
\caption{
Gap functions for the parameter values of Fig.\ref{fig14}(a).
\label{fig15}}
\end{center}
\end{figure}
\begin{figure}[htb]
\begin{center}
\scalebox{0.8}{
\includegraphics[width=10cm,clip]{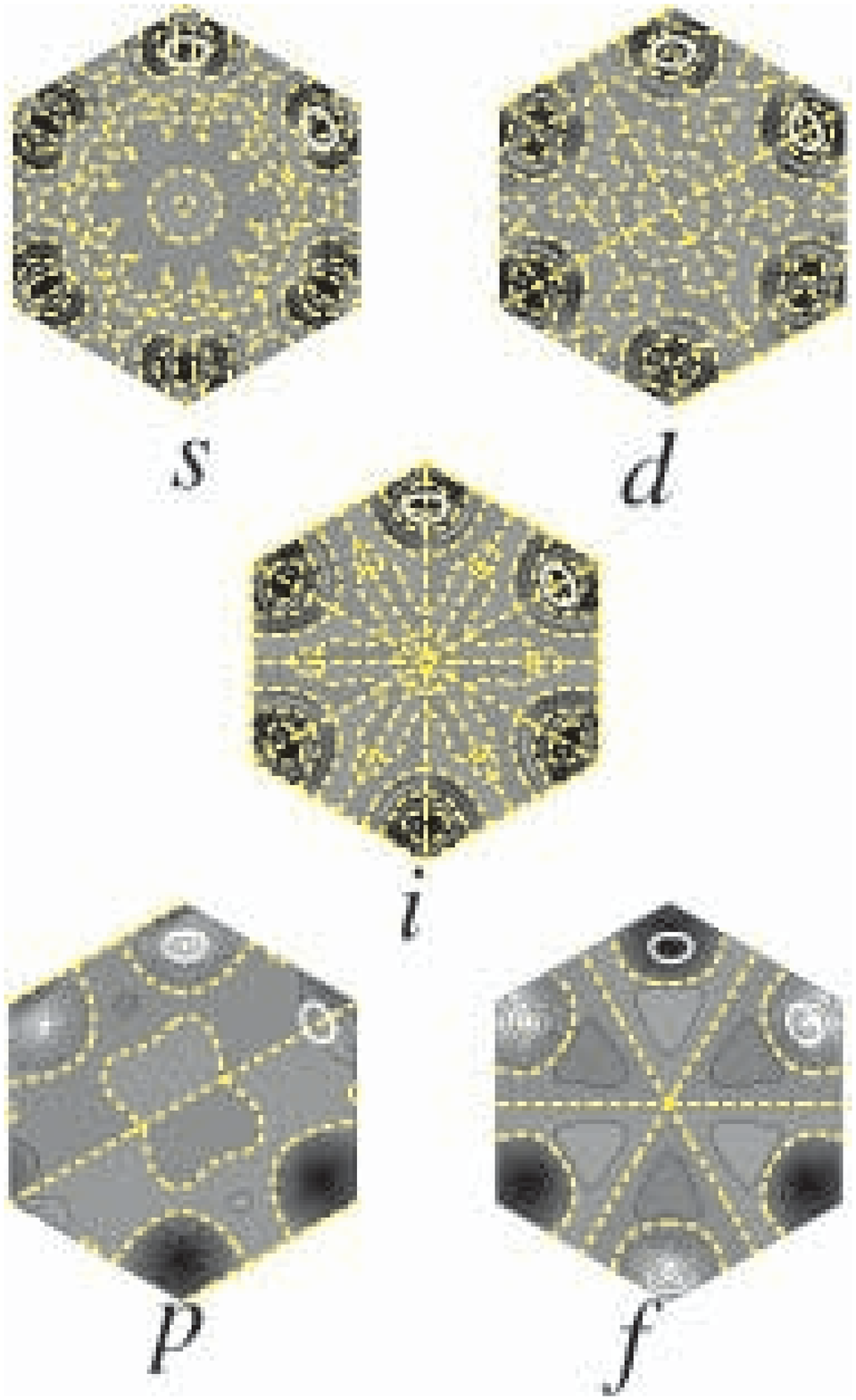}}
\caption{
Gap functions for the parameter values of Fig.\ref{fig14}(b).
\label{fig16}}
\end{center}
\end{figure}

As shown in Fig.\ref{fig14}(a), 
the triplet pairings dominate strongly over singlet pairings,
and among the triplet pairings, $p$-wave has the largest $T_c$. 
This result can again be understood from the above real space picture. 
Namely, since the spin susceptibility 
peaks around the $\Gamma$ point, the wave length of the spin alignment is 
large, so that the singlet pairing distance is large from 
Fig.\ref{fig12}. In fact, the 
number of nodes in the singlet pairing gaps are found to increase 
compared to those in section\ref{large}, as seen in Fig.\ref{fig15}. 
This makes the singlet pairing 
less competitive against triplet pairings.
Moreover, since the wave length of the spin alignment is large, not only 
electrons residing at nearest neighbors but also those separated 
to some extent have parallel spins, so that spin triplet pairing
can take place at relatively large distances, 
thereby circumventing the effect of 
off-site repulsions. It can be seen from Fig.\ref{fig15}
that the $f$-wave gap has additional nodes close to the Fermi surfaces 
suggesting that the pairing takes place not only at nearest neighbors 
but also at larger distances. Pairing at large distances, and thus 
the presence of nodes near the Fermi surfaces, degrades $f$-wave 
pairing, so that $p$-wave pairing dominates.

Even if we decrease $V_1$ down to $V_1=0.7$, $p$-wave still dominates 
over $f$-wave as seen Fig.\ref{fig14}(b), in contrast to the case 
discussed in sections \ref{moderate}. This is because 
in the presence of purely ferromagnetic spin fluctuations, 
$f$-wave pairing takes place at large distances 
even for relatively small off-site repulsions, 
so that the additional nodes are located close to 
the Fermi surfaces. This can in fact be seen by comparing Fig.\ref{fig15} and 
Fig.\ref{fig16}, where the additional nodes in the 
$f$-wave gap do not move away so much from the Fermi surfaces even if we 
reduce $V_1$ from 1.0 to 0.7. This is in contrast with the case of 
Fig.\ref{fig7} and Fig.\ref{fig12}, where the additional nodes 
in the $f$-wave gap clearly moves away 
from the Fermi surfaces (i.e., the nearest neighbor pairing component 
becomes large)  as we reduce $V_1$.

\subsection{Peculiar Disconnectivity of the Fermi surfaces}
The reason why $i$-wave pairing, which has a very large angular momentum,
may dominate in the presence of off-site repulsions 
can be found in the disconnectivity of the 
Fermi surface {\it peculiar} to the present system. 
Namely, the Fermi surfaces in the present case are 
disconnected in a way that they do not intersect the 
$\Gamma$-M lines nor the hexagonal Brillouin zone edge, 
both of which are the nodal lines of the $i$-wave gap. 
Thus, the number of nodal lines that intersect the 
Fermi surfaces is similar among the three singlet 
pairing symmetries, extended $s$, $d$, and $i$-waves,  as seen in 
Fig.\ref{fig11}, thereby making the competition subtle.
In fact, we have looked into various sets of parameter values and 
found that the three singlet symmetries tend to have close values of 
$\lambda$.
\begin{figure}[htb]
\begin{center}
\scalebox{0.8}{
\includegraphics[width=10cm,clip]{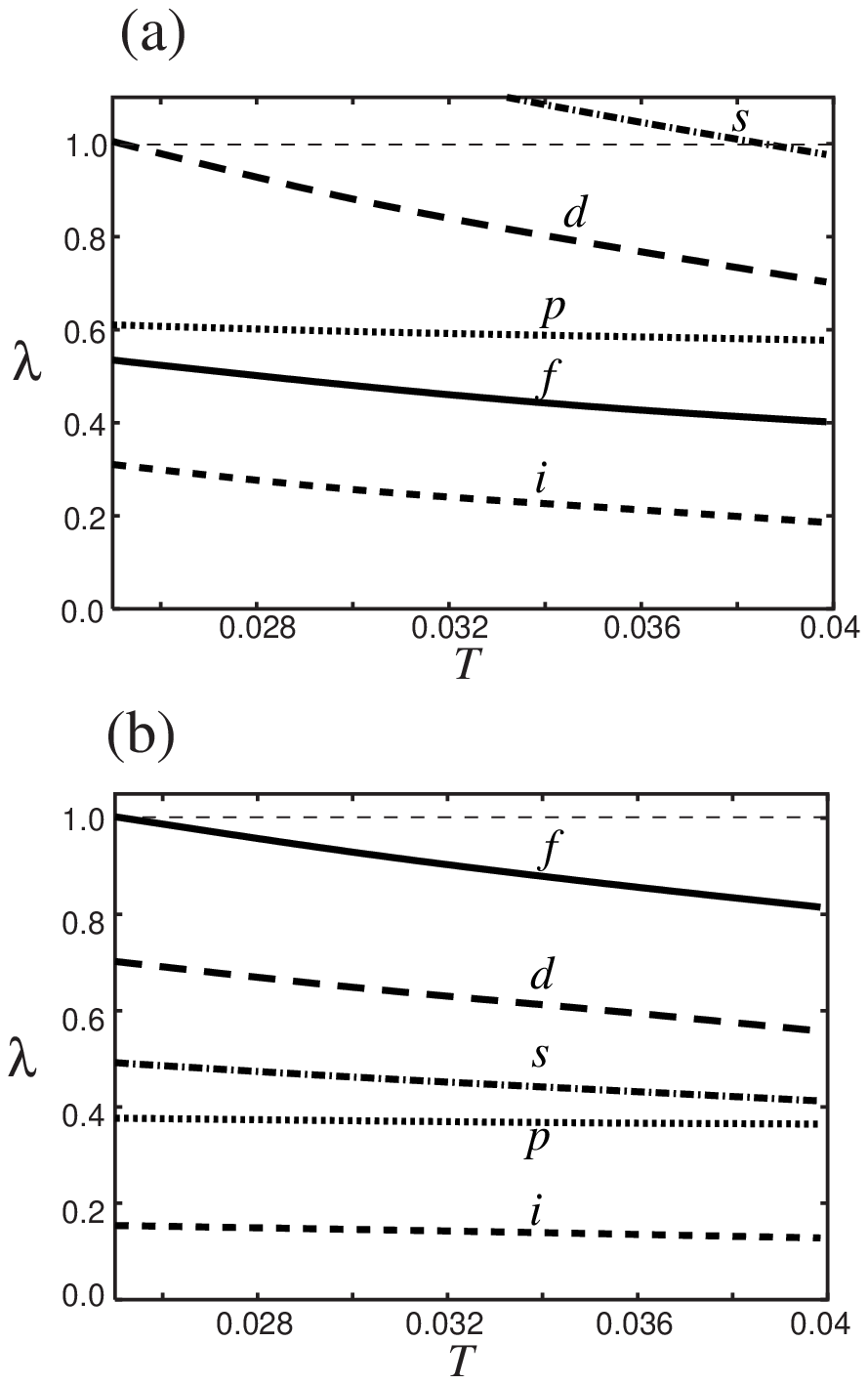}}
\caption{
Plots similar to Fig.\ref{fig6} for  $t_2=t_3=t_4=0$. 
$U=2.4$, $V_1=1.0$, $V_2=0.4$, and $n_h=0.4$ (a), 
$U=3.6$, $V_1=1.5$, $V_2=0.6$, and $n_h=0.18$ (b).
\label{fig17}}
\end{center}
\end{figure}
\begin{figure}[htb]
\begin{center}
\scalebox{0.8}{
\includegraphics[width=10cm,clip]{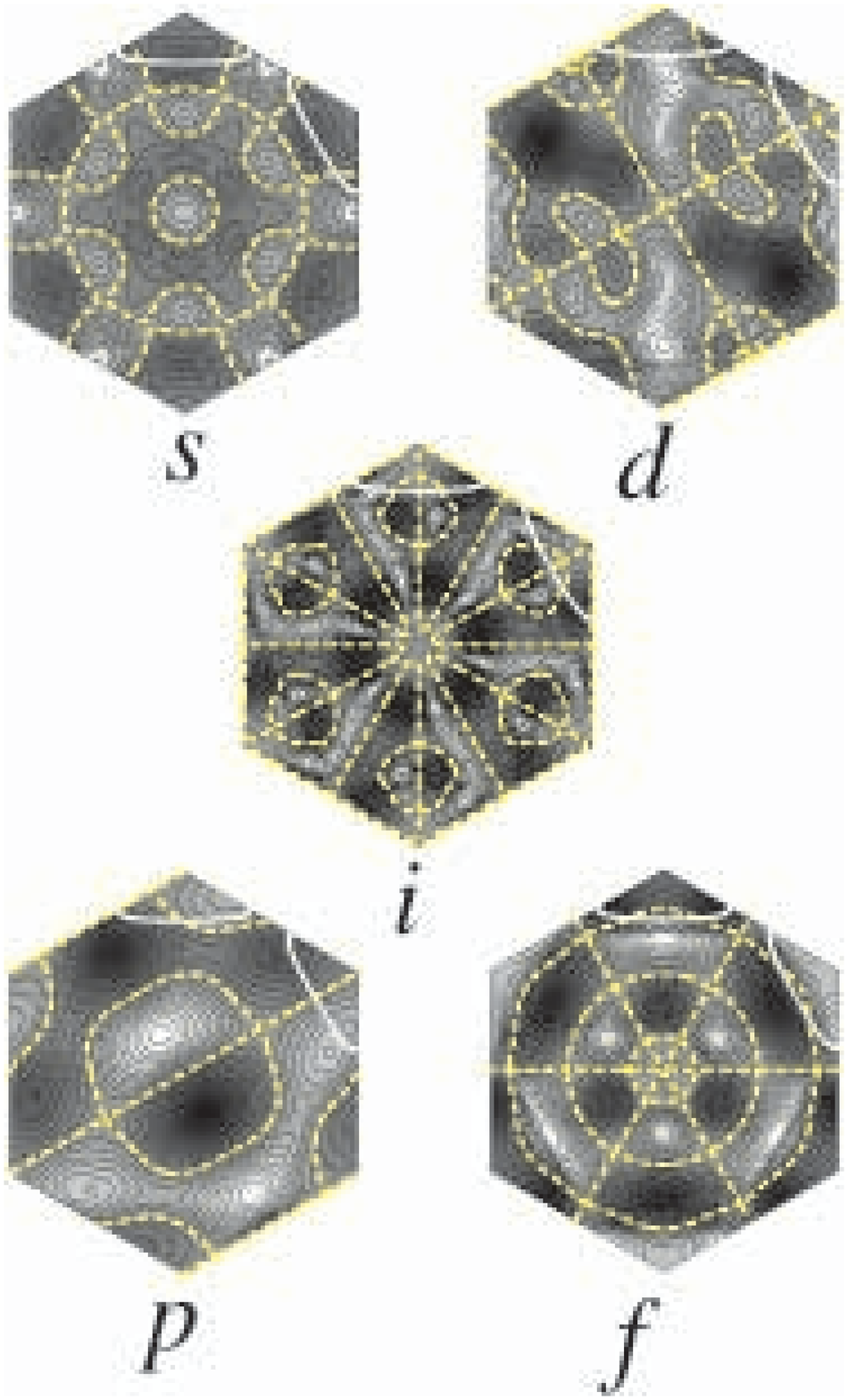}}
\caption{
Gap functions for the parameter values of Fig.\ref{fig17}(a).
\label{fig18}}
\end{center}
\end{figure}
\begin{figure}[htb]
\begin{center}
\scalebox{0.8}{
\includegraphics[width=10cm,clip]{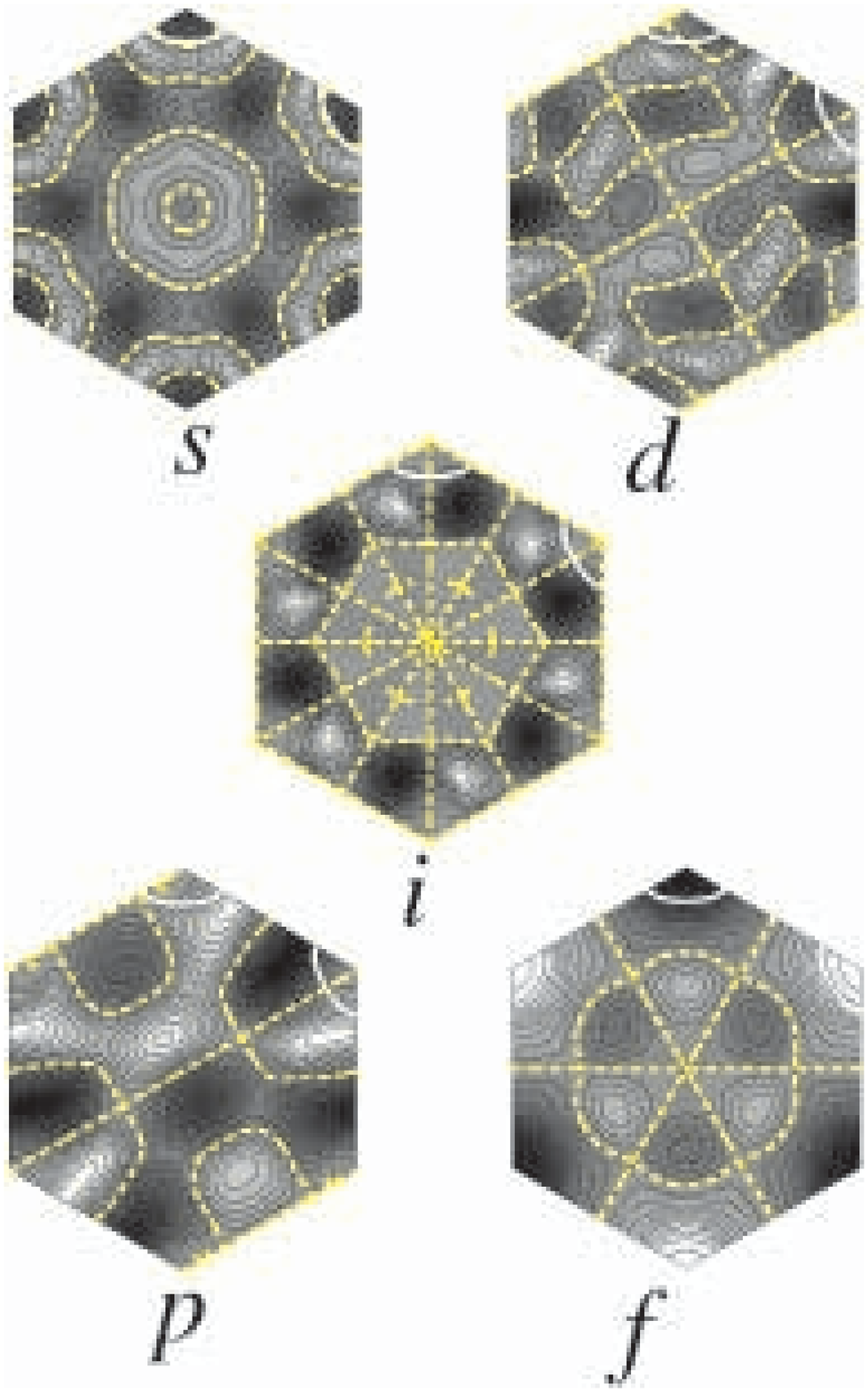}}
\caption{
Gap functions for the parameter values of Fig.\ref{fig17}(b).
\label{fig19}}
\end{center}
\end{figure}

To show more clearly that the subtle competition between $i$-wave and other 
singlet pairings is due to the disconnectivity of the 
Fermi surfaces peculiar to the present system, we compare the above results 
with those for cases with only nearest neighbor hopping, 
namely with a simple triangular lattice. In a simple triangular lattice
with $n_h<0.5$, the Fermi surfaces are disconnected 
in a manner that they do not intersect the $\Gamma$-M lines,
but they do intersect the Brillouin zone edge. 
The absence of $t_4$ 
moves the van Hove singularity away from the band top, so 
in order to have a moderate $T_c$ as in the cases above,
we either have to take larger values of $n_h$, or increase the values of the 
electron-electron interactions. Here we take 
$n_h=0.4$ and the interactions the same as in section\ref{large}
in Fig.\ref{fig17}(a), while in Fig.\ref{fig17}(b) we take the same $n_h$ but 
take the interactions 1.5 times larger than 
in section\ref{large}. It can be seen that in 
both cases, $\lambda$ of the $i$-wave pairing is much smaller than that of 
the $d$- and the $s$-wave pairings. This is because 
the Fermi surfaces intersect the Brillouin zone edge, i.e., the nodes of 
the $i$-wave gap, as seen in Fig.\ref{fig18} and Fig.\ref{fig19}.

\subsection{Comparison with the case with antiferromagnetic spin fluctuations}
In the present study, we have looked into a model where 
nearly ferromagnetic spin fluctuations and charge fluctuations 
due to off-site repulsions coexist. 
It is interesting to compare the present situation with the case 
where {\it antiferromagnetic} 
spin fluctuations and charge fluctuations coexist. 
We have previously 
proposed that spin-triplet pairing is possible in a  
quarter-filled, quasi-1D organic material (TMTSF)$_2$X,\cite{KAA,KYTTA}
where superconductivity lies 
right next to spin density wave (SDW) 
in the pressure-temperature phase diagram, so 
that the spin fluctuations should be antiferromagnetic.
Naively, antiferromagnetic spin fluctuations would favor 
spin singlet $d$-wave pairing, while experiments suggest triplet 
pairing.\cite{Lee1,Lee2} 
Our proposal is that if $2k_F$ ($=\pi/2$ because the band is 
quarter-filled) charge fluctuations coexist
with $2k_F$ spin fluctuations, as suggested from an experimental
fact that CDW actually coexist with SDW\cite{Pouget,Kagoshima}, 
spin triplet `$f$-wave' like pairing, which 
is essentially a fourth neighbor pairing having a gap form of $\sin(4k_x)$ 
(this is called $f$-wave in the sense that the gap changes sign 
as $+-+-+-$ along the Fermi surface),
 may become competitive against singlet `$d$-wave' like pairing, 
which is a second neighbor pairing with a gap form $\cos(2k_x)$. 
In momentum space, this close competition arises because 
(i) the number of nodes of the 
$f$-wave gap on the Fermi surface is the same with that of the 
$d$-wave because of the disconnectivity of the Fermi surfaces
due to quasi-one-dimensionality, and (ii) the pairing interactions 
due to spin and charge fluctuations (apart from the first order terms) 
have the form 
$V^s=\frac{3}{2}V_{\rm sp}-\frac{1}{2}V_{\rm ch}$ for singlet 
pairing and $V^t=-\frac{1}{2}V_{\rm sp}-\frac{1}{2}V_{\rm ch}$ for 
triplet pairing, where $V_{\rm sp}$ and $V_{\rm ch}$ are 
contributions from spin and charge fluctuations, respectively, 
so that the absolute values of $V_s$ and $V_t$ becomes 
comparable when $V_{\rm sp}\simeq V_{\rm ch}$.
A similar theory has been proposed by Fuseya {\it et al.}\cite{Fuseya}
Recently, we have confirmed this scenario microscopically
by applying RPA to an extended Hubbard model with off-site 
repulsions up to third nearest neighbors.\cite{YTK}
Here, a key parameter is the second nearest neighbor repulsion,
due to which the $2k_F$ charge fluctuations arises.

\begin{figure}[htb]
\begin{center}
\scalebox{0.8}{
\includegraphics[width=10cm,clip]{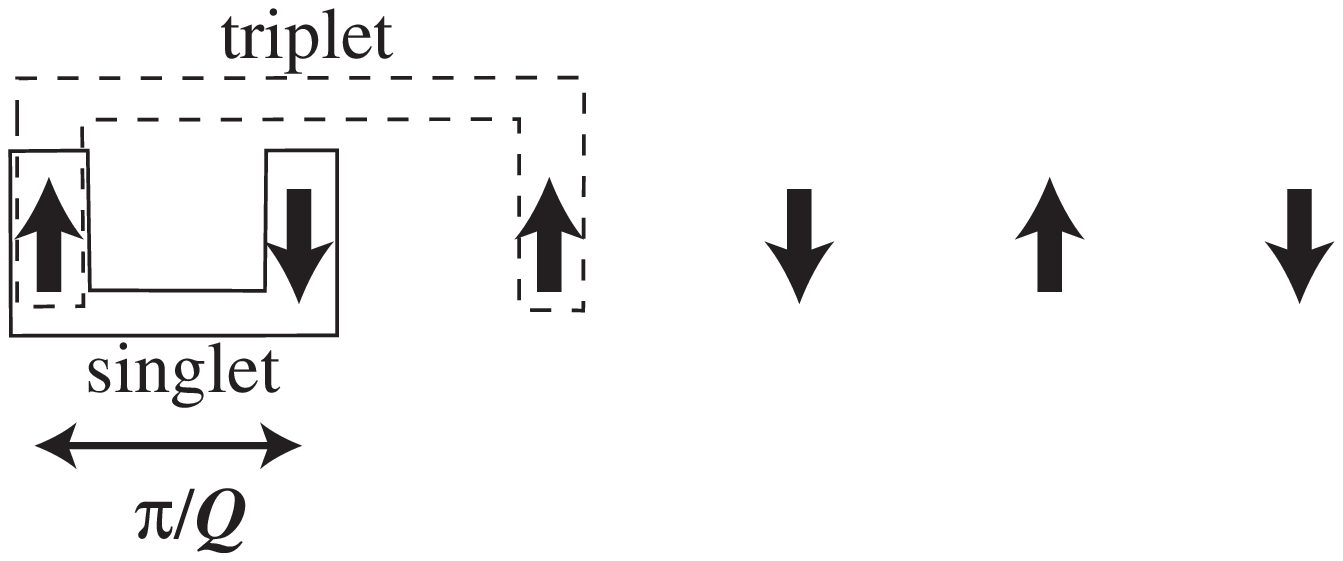}}
\caption{
An intuitive real space picture of singlet and triplet pairings 
when antiferromagnetic spin fluctuations are present.
Note that when $Q=2k_F=\pi/2$ as in the case of (TMTSF)$_2$X, 
arrows in this figure are separated by two lattice spacings.
\label{fig20}}
\end{center}
\end{figure}

In real space, this $f$-$d$ competition can be understood as 
follows. When there are purely antiferromagnetic spin fluctuations,
second neighbor singlet pairing is likely to occur, as seen 
from Fig.\ref{fig20}. If we now turn on the second nearest neighbor 
repulsion, the second neighbor singlet pairing is degraded,
and a fourth neighbor triplet pairing becomes competitive.
Since this intuitive picture in real space is rather general,
the tendency that spin singlet pairing, which is favored by 
antiferromagnetic spin fluctuations, gives way to triplet 
pairings in the presence of sufficiently large  
off-site repulsions, should hold in various situations.
In fact, such a singlet-triplet crossover due to off-site repulsions is 
seen also in a half-filled quasi-1D system,\cite{Onari}
a model for Sr$_2$RuO$_4$ on a 2D square lattice,\cite{Arita}
and a model for Na$_x$CoO$_2$ on a 2D triangular lattice.
\cite{TanaOgata}

Now, it is interesting to point out that 
the above tendency is exactly the opposite of what has happened in the 
present case with nearly ferromagnetic spin fluctuations : 
triplet pairings formed at relatively short distances  give way to 
singlet pairings formed at farther distances when 
sufficiently large off-site repulsions are introduced.
Thus, the effect of the off-site repulsion totally 
differs depending on whether the spin fluctuations are 
ferromagnetic or antiferromagnetic.
It is also important to note that in both cases, 
the disconnectivity of the 
Fermi surfaces can assist the pairings at large distances 
(singlet in the ferromagnetic case and 
triplet in the antiferromagnetic case), 
which have larger number of nodes in the gap function.

\subsection{Validity of focusing on the pockets}
In the present study, we have focused only on the $e_g'$ Fermi 
pockets. 
Although it is necessary to consider a multiband model\cite{Koshibae} 
in order to strictly take into account the effect of the Fermi pockets, 
our single band approach is supported by   
a recent multiband FLEX study,\cite{MYO}
where a similar $f$-wave pairing is found to dominate when 
spin fluctuations are not too purely ferromagnetic.

Experimentally, 
the Fermi pockets are not observed in ARPES experiments
\cite{Hasan,Yang} up to date.
However, since these experiments are done for materials with 
relatively large Na content,
(i.e., large number of electrons in CoO layers) 
it is likely that the Fermi level lies above the $e_g'$ bands.
\cite{Karppinen} In fact, an experimental result 
that maximum $T_c$ is reached only when the content of 
Na decreases sufficiently\cite{Schaak} can be considered as 
an indirect support for scenario in which the $e_g'$ 
band plays an important role for superconductivity.

\subsection{Pairing symmetry in the cobaltate}
Let us finally discuss the pairing symmetry in the cobaltate in  
light of the present results along with the existing experiments.
We have seen that ``full-gapped'' 
$f$-wave proposed in ref.\onlinecite{KYTA} is robust 
to some extent even in the presence of off-site repulsions.
Existence of nodes in the gap function is suggested in a
number of experiments,\cite{Fujimoto,Ishida}
which may seem to contradict with the $f$-wave scenario.
In the actual Na$_x$CoO$_2$, however,
there is also the large Fermi surface around
the $\Gamma$ point, which is not taken into account in the present study.
Then we can propose a possible scenario where the `active' Fermi surfaces for 
superconductivity are the pockets, while the large Fermi surface is 
a 'passive' one, in which superconductivity is 
induced by the `active' band via interband interactions.
Since the nodes of the $f$-wave gap intersects the 
large Fermi surface, this can account for the experimental results 
suggesting the existence of gap nodes. In fact, it has 
been shown that a $\mu$SR data can be 
explained by taking this view.\cite{Uemura}

On the other hand, singlet pairings are found to 
dominate over triplet pairings in the present model 
when spin fluctuations are not purely ferromagnetic and 
the off-site repulsions are sufficiently large.
Among the singlet symmetries, $i$-wave is another good candidate
that competes with $f$-wave since it does not break time
reversal symmetry below $T_c$. We have shown that $i$-wave 
may dominate over other symmetries, but the 
parameter range where this symmetry dominates 
is not so wide as that for the $f$-wave.
Although extended $s$-wave is also consistent with almost all the 
existing experiments, we could not find a parameter regime where 
this symmetry dominates. However, considering the fact that 
the competition within the singlet pairings is very subtle, and that 
the present model only roughly takes into account the 
electronic structure of the 
actual material, we believe the 
possibility of extended $s$-wave pairing cannot be ruled out. 

Since $d$-wave and $p$-wave pairings 
are likely to result in a broken time reversal symmetry 
and/or a full gapped state (unless an accidental situation occurs)  
below $T_c$ due to the two-fold degeneracy,  
we believe that the possibility of these pairing states are less likely.
We may conversely say from the present results that the 
spin fluctuation in the actual cobaltate is not {\it purely} 
ferromagnetic and/or the off-site repulsions are not too large,
so that $p$-wave pairing does not dominate.

At present, it is not clear experimentally whether the 
pairing occurs in the singlet or in the triplet channel 
because Knight shift experiments are done on polycrystal samples.
In order to settle this singlet-triplet debate, 
Knight shift experiment for high quality samples, or 
alternatively, tunneling spectroscopy studies based on a newly 
developed theory for triplet superconductors 
proposed by one of the present authors\cite{TanakaKas}
is required. Thus, our viewpoint is that possibility of neither 
singlet nor triplet pairings cannot be ruled out experimentally 
at the present stage. 
Then, the bottom line obtained from the present study, 
combined with the existing 
experimental facts, is that triplet $f$-wave 
pairing is most likely, although there remains a 
possibility of singlet $i$-wave pairing, 
and also a possibility of extended $s$-wave pairing cannot be ruled out.
This ambiguity is a direct consequence of the very 
point of the present study. 
Namely, the present system is highly peculiar in that anisotropic 
pairings  having large angular momentum, such as $f$- and $i$- waves, are 
good candidates due to the disconnectivity of the Fermi surfaces. 
The fact that these ``unusual'' pairing symmetries are 
good candidates makes the pairing competition subtle.

\section{Summary}
To summarize, we have studied an effective single band model 
for the hydrated sodium cobaltate focusing on the $e_g'$ hole pockets, 
where we take into account 
 off-site repulsions up to second nearest 
neighbors. We find that $f$-wave pairing proposed in our 
previous study is robust to some extent in the presence of 
off-site repulsions. For large off-site repulsions, $f$-wave 
gives way to singlet pairings 
that does not break time reversal symmetry below $T_c$. 
\acknowledgements
Calculation has been performed
at the facilities of the Supercomputer Center,
Institute for Solid State Physics,
University of Tokyo
%======Reference===================================
%

%===============================================================


\begin{thebibliography}{99}
\bibitem{comment} $t_2$, $t_3$, $t_4$ 
should be considered as fitting parameters
determined so as to reproduce the upper portion of the $e_g'$ bands in 
momentum space, rather than the amplitude of hopping from 
one $3d$ orbital to another.
\bibitem{Karppinen} This view may further be supported by a recent study
by Karppinen {\it et al.} (cond-mat/0402484),
which shows that the actual Co valence can be smaller (the Fermi level can be 
higher) than expected from the composition Na$_x$CoO$_2$.
\end{thebibliography}
\end{document}